\documentclass[a4paper,11pt]{article}
\usepackage{pos}
\usepackage[export]{adjustbox}

\title{Monitoring Gamma Ray Bursts with the Zirè instrument onboard the NUSES
space mission}

\author*[a,b]{R. Sarkar}
\onbehalf{on behalf of the NUSES collaboration}

\affiliation[a]{Gran Sasso Science Institute, Via Iacobucci 2, I-67100,
L’Aquila, Italy}
\affiliation[b]{Istituto Nazionale di Fisica Nucleare --- Laboratori Nazionali
del Gran Sasso, Via G. Acitelli 22, I-67100 Assergi, L’Aquila, Italy}

\emailAdd{ritabrata.sarkar@gssi.it}

\abstract{
 The Zirè experiment onboard NUSES space mission has several science goals,
 including the measurements of charged particles and light nuclei from few up to
 hundreds of MeVs, for the study of low-energy cosmic rays, space weather
 phenomena, and possible magnetosphere-lithosphere-ionosphere coupling signals.
 Furthermore, the experiment intends to test new tools for the detection of
 photons in the energy range of about 0.03--50~MeV, allowing the investigation of
 transient phenomena like gamma-ray bursts (GRBs). A high-density segmented
 calorimeter exploiting novel scintillator crystals and silicon photomultiplier
 technology for the readout system will serve this purpose along its main task
 of calorimetric energy measurement of the cosmic-ray charged particles. In
 this work, we discuss the functionality of the calorimeter as a GRB monitor
 through the calculation of its performance in terms of effective area,
 sensitivity, and timing response for the transient outbursts.
 }

\ConferenceLogo{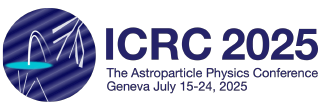}

\FullConference{39th International Cosmic Ray Conference (ICRC2025)\\
 15–24 July 2025\\
Geneva, Switzerland\\}

\begin{document}
\maketitle

\section{Introduction}
\label{sec:intro}

The NUSES space mission \cite{DeMitri_2023}, proposed by the Gran Sasso Science
Institute (GSSI) in collaboration with the Istituto Nazionale di Fisica Nucleare
(INFN) and by Thales Alenia Space Italy (TAS-I) involving many other institutes
and universities from Europe and the US, consists of two main instruments:
Terzina and Zirè. The NUSES spacecraft will fly on a low-Earth orbit (LEO) at an
altitude of 550~km at a high inclination of 97.8$^\circ$, traveling in a
Sun-synchronous mode along the day/night boundary line.

Terzina is devoted to the exploration of new observational approaches in the
study of ultra-high-energy cosmic rays and neutrino astronomy. The primary goals
for Zirè \cite{Aloisio_2023SF} are to measure electrons, protons, and light
nuclei from a few up to hundreds of MeVs, study of low-energy cosmic rays (CRs),
space weather phenomena, and possible magnetosphere-lithosphere-ionosphere
coupling signals. This will investigate possible correlations between natural
activities occurring on Earth and unusual transient phenomena in the surrounding
magnetosphere and ionosphere, such as particle bursts. A further goal of the
experiment is to test new techniques for the detection of photons in the energy range of about 0.03--50~MeV based on silicon photomultiplier (SiPM) technology.
This will allow for the investigation of transient phenomena, such as Gamma
Ray Bursts (GRBs), electromagnetic follow-up of gravitational wave events,
supernova emission lines, etc., and also study persistent $\gamma$ ray sources. In complement with the other instruments onboard Zirè, the possible correlations
of bright GRBs with local effects on the charge particle radiation environment
might also be studied \cite{Battiston_2023}. The technical knowledge gained from
this mission as a pathfinder will be used for future space missions for transient
$\gamma$-ray source detection, like Crystal~Eye \cite{Aloisio_2025103171}.

\section{The Zirè detector}
\label{sec:detDesc}

\begin{figure}
  \centering
  \includegraphics [width=0.5\columnwidth]{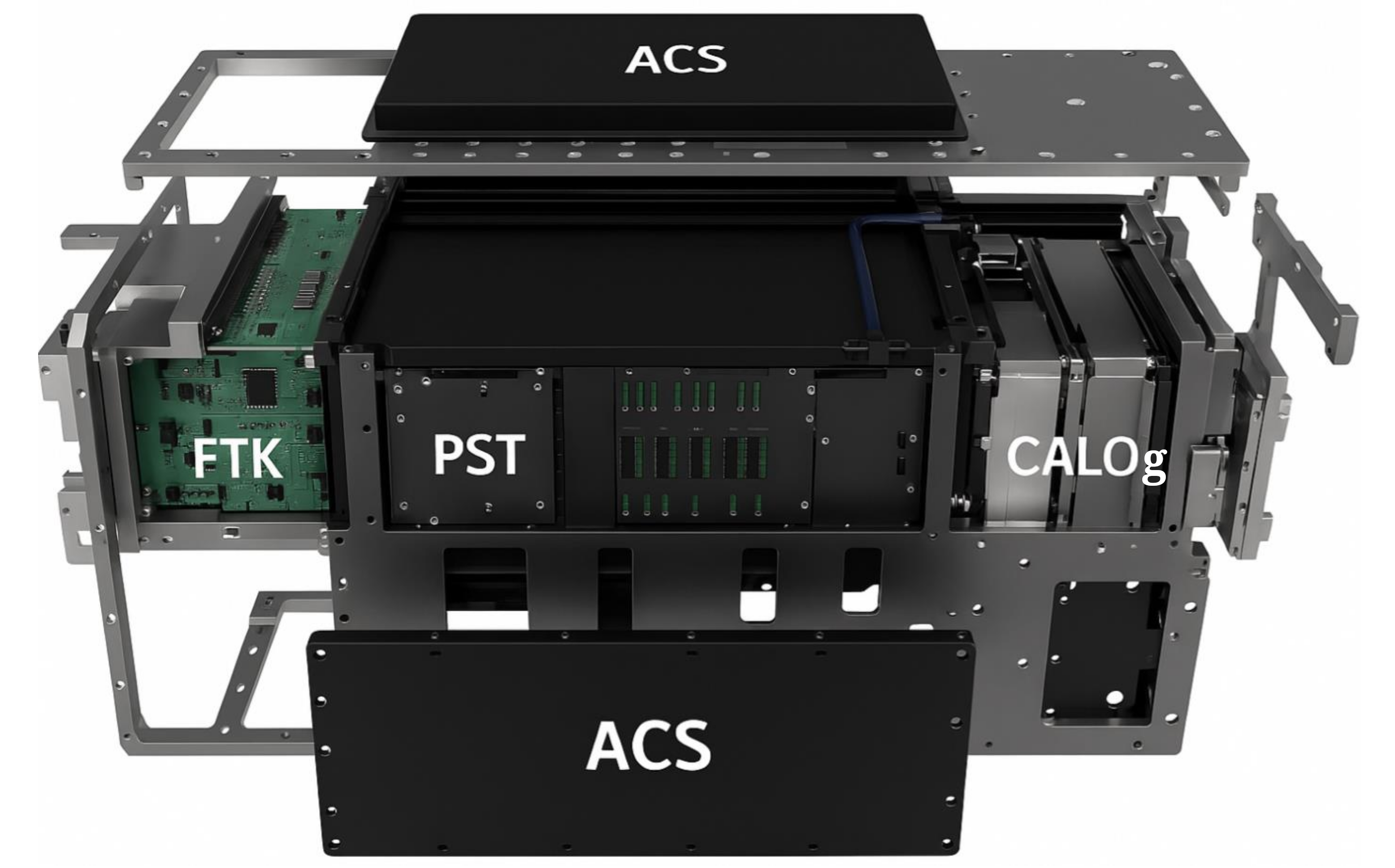}
  \caption{Layout of the Zirè payload. Particles entering from the left window
  will go through FTK, PST, and CALOg. The ACS surrounds the detector. Two more
  thin windows (not shown here) are open on the two sides of the CALOg to
  directly get low-energy photons for transient source detection.}
  \label{fig:zire}
\end{figure}

A schematic view of the computer-aided design (CAD) for the Zirè detector module
is shown in Fig.~\ref{fig:zire}. Starting from the left in the figure, the
detector comprises the following subdetectors.

\begin{itemize}
  \item {\bf Fiber TracKer (FTK):} FTK contains three X-Y modules with a cross  section of 9.6~$\times$~9.6~cm$^2$ and 2.5~cm spacing in between. Each module consists of two layers with mutually orthogonal fibers. Fibers have a double layer structure consisting of a polystyrene core (inner side) with a fluorescent agent (n = 1.59), an inner cladding of polymethylmethacrylate (PMMA) (n = 1.49), and an outer cladding of fluorinated polymer (n = 1.42). The total thickness of each module is only about 2~$\times$~0.8~mm to reduce multiple Coulomb scattering and to keep the energy loss of the charged particles as low as possible.

  \item {\bf Plastic Scintillator Tower (PST):} A tower of 32 plastic scintillator (PS) layers, each composed of three PS X-Y bars. The first six layers of the PST, namely the ones close to the FTK, are of 12~$\times$~12~$\times$~1~cm$^3$, while the other 26 layers have dimensions of 12~$\times$~12~$\times$~0.5~cm$^3$;

  \item {\bf Calorimeter (CALOg):} The granular calorimeter is made by a matrix of 4~$\times$~4~$\times$~2 gadolinium aluminum gallium garnet (GAGG) cuboids with size 2.5~$\times$~2.5~$\times$~3.0~cm$^3$;

  \item {\bf Anti-Coincidence System (ACS):} 9 PS layers with 0.5~cm of thickness working as the veto detector surrounding the instrument from all the directions except for the one where the FTK is located.
\end{itemize}

The readout of each subdetector will be performed with SiPMs, replacing the
photomultiplier tubes mostly used for the experiments in space. The innovative
use of SiPMs for both the Zirè and Terzina experiments onboard the NUSES
satellite is one of the features of this space mission. A full SiPM-based
technology turns out to be the best choice not only for the reduced size and
low power-consumption of such sensors but also for their excellent detection
capability.

\section{Effective area}
\label{sec:effarea}

\begin{figure}
  \centering
  \includegraphics [width=0.25\columnwidth,margin=0pt 0pt,valign=c]{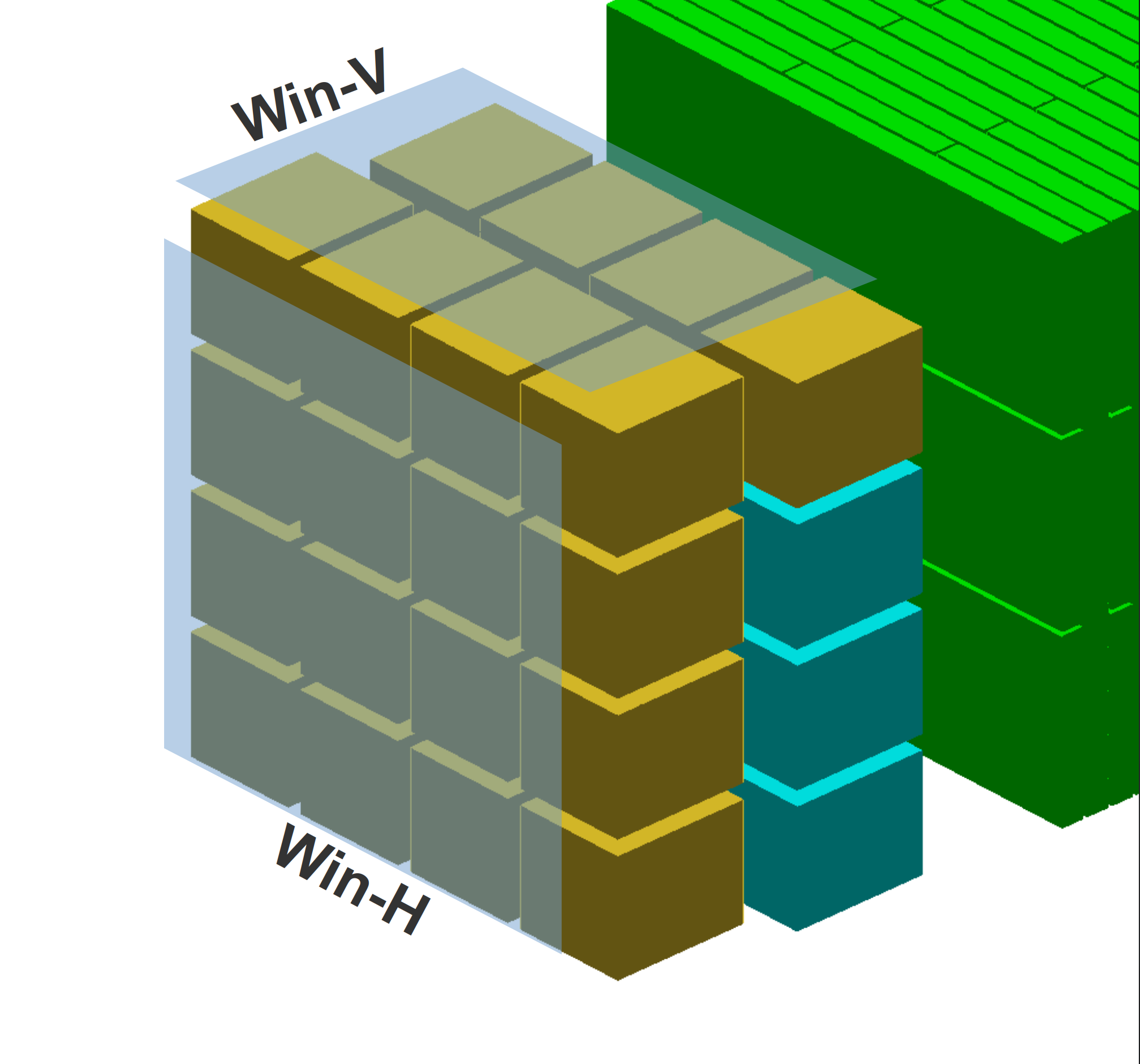}
  \includegraphics [width=0.50\columnwidth,margin=0pt 0pt,valign=c]{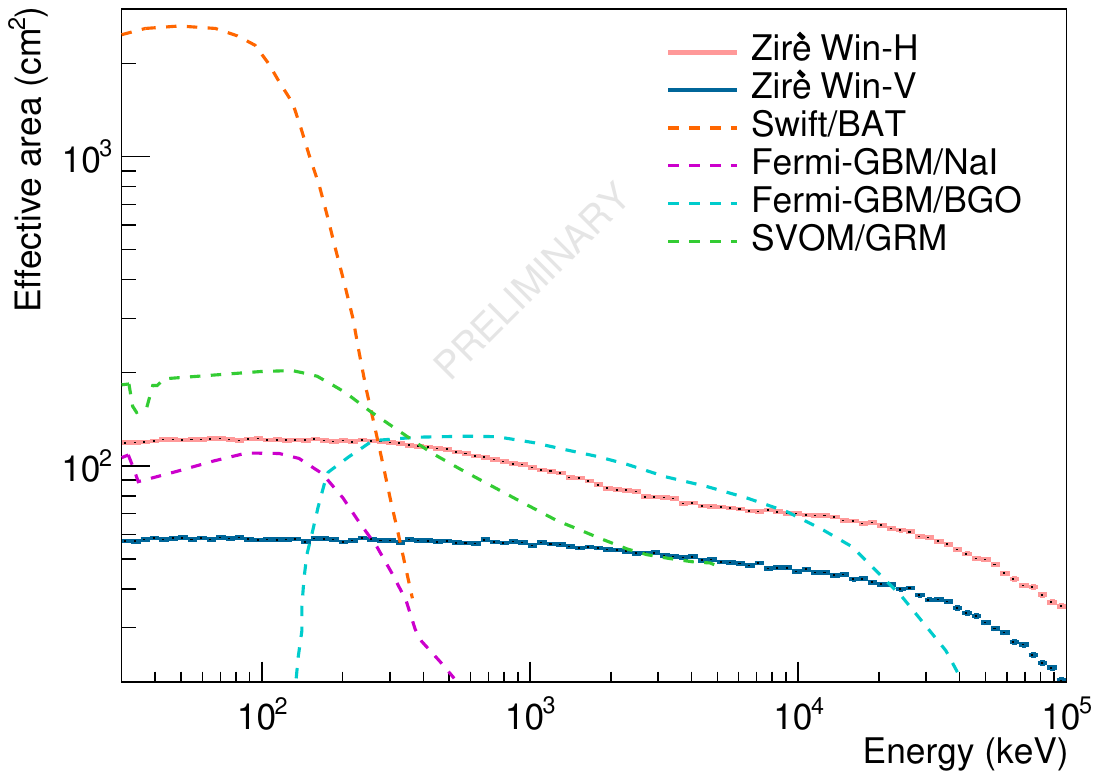}
  \caption{(Left:) Position of the two windows of the Zirè CALOg for direct
  GRB detection. (Right:) CALOg effective area for observation through the two
  windows. Effective area for other GRB instruments in the similar range such as
  Swift/BAT \cite{barthelmy_2005}, Fermi-GBM (NaI and BGO) \cite{meegan_2009},
  and SVOM/GRM \cite{he_2025} are also shown for comparison.}
  \label{fig:effarea}
\end{figure}

The CALOg will also be used independently for the study of low-energy 
$\gamma$-rays in the energy range between 0.03--50~MeV. Two windows in the
structure surrounding the CALOg have been specifically included in the design
for this purpose (see left panel of Fig.~\ref{fig:effarea}), one pointing to the
zenith (Win-V) and another to the horizon (Win-H). 

We calculated the effective area of CALOg in both windows using the Geant4
simulation of the detailed geometrical model of the detector. To select
high-quality events, maximize background rejection, and avoid electronic noise,
we applied some threshold cuts for event signals (7~keV and 30~keV for each ACS
tile and CALOg crystal, respectively). On the top, we requested the total
deposited energy in the ACS layer to be $<$ 300~keV, to select photons over
charged-particle events. The right panel of Fig.~\ref{fig:effarea} shows a
preliminary estimate of the effective area of CALOg through either window, along
with the effective area of other GRB instruments for comparison.

\section{Detector background}
\label{sec:bkg}

As already mentioned, the NUSES space mission is going to operate in a
Sun-synchronous LEO at an altitude of about 550~km and with a high inclination
of 97.8$^\circ$. Due to the high inclination of its orbit, the detector will
undergo a diverse radiation environment from the equatorial region to the polar
region. In this environment, background radiation includes primary particles
of cosmic and solar origin and secondary products from the interaction of
high-energy cosmic particles with the Earth atmosphere. These include cosmic
diffuse $\gamma$-ray photons; albedo X-ray and $\gamma$-ray photons from the
Earth atmosphere; albedo neutrons; primary and trapped protons; primary and
trapped e$^-$ and e$^+$. We used the flux distribution of different background
components at LEO compiled by \cite{cumani_2019}, to predict the background
environment at the operational orbit of Zirè, while using a moderate solar
modulation potential (650~MV). 

\begin{figure}
  \centering
  \includegraphics [width=0.45\columnwidth]{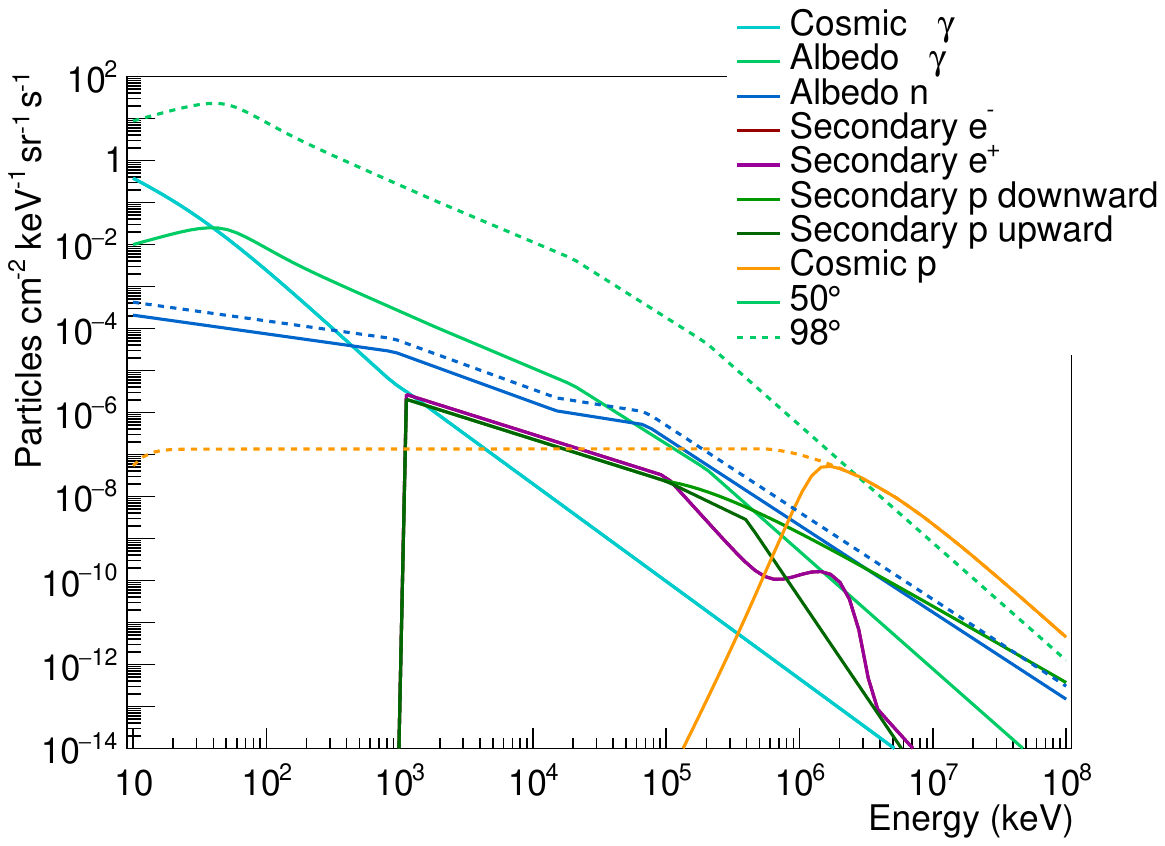}
  \includegraphics [width=0.45\columnwidth]{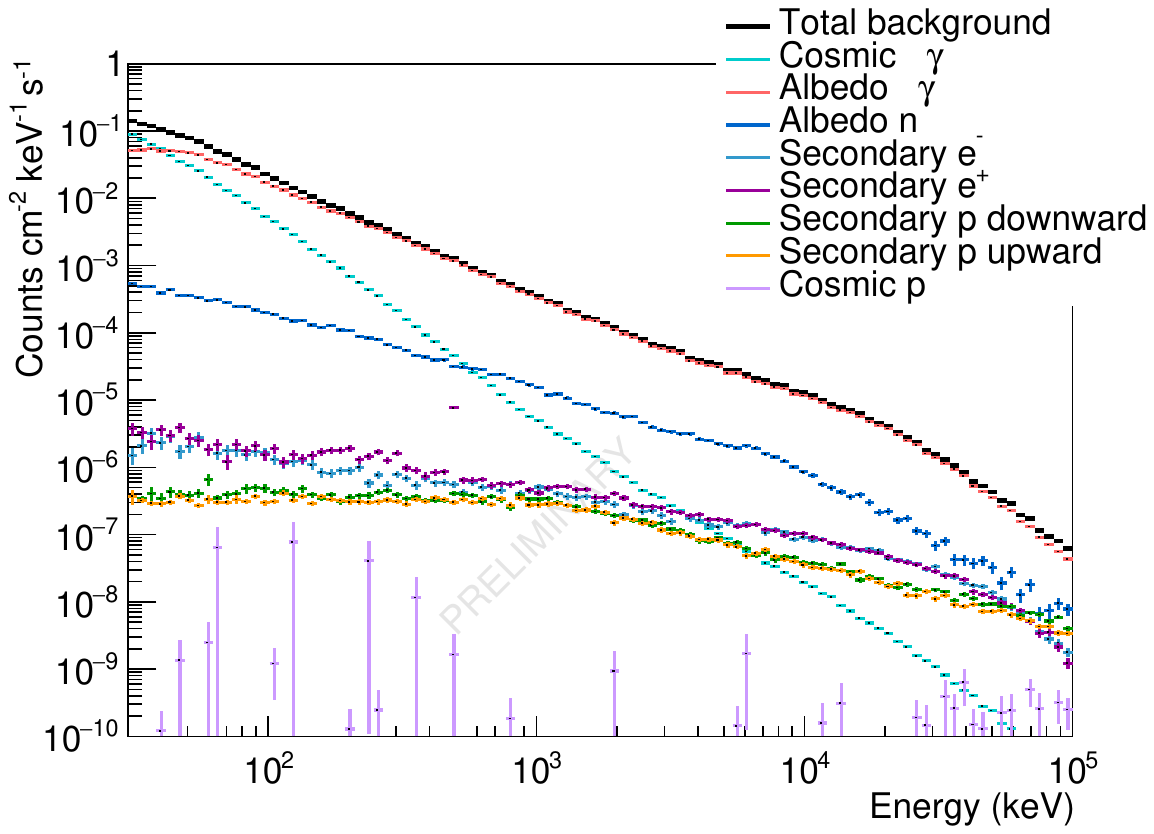}
  \caption{(Left:) In-orbit background radiation fluxes for different contributing components at 50$^\circ$ (solid lines) and 98$^\circ$ (dashed
  lines) latitudes. (Right:) Effective background component fluxes in CALOg for
  the radiation environment at 50$^\circ$ latitude.}
  \label{fig:bkg}
\end{figure}

We simulated the interactions of the background components with the detector mass
model to calculate the background induced in the detector. For this purpose, we
used radiation environment information at three different geomagnetic latitudes
of 5$^\circ$, 50$^\circ$, and 98$^\circ$. In the left panel of Fig.~\ref{fig:bkg}
we show the fluxes from the different components of the radiation environment at
50$^\circ$ and 98$^\circ$, while the contribution at 5$^\circ$ is even lower than
50$^\circ$ and not shown here. The partial and total contributions of the
detector background due to different components of the external radiation and
particles (at 50$^\circ$) are shown in the right panel of Fig.~\ref{fig:bkg}.
It is clear that, for the trigger condition mentioned in
Section~\ref{sec:effarea}, the dominant background contribution comes from
$\gamma$ rays, both from the cosmic diffuse and atmospheric components
mentioned above.

\section{Sensitivity to transient sources}
\label{sec:sens}

We calculated the sensitivity of the detector for short-duration transient
sources such as GRBs. The signal-to-noise ratio (S/N) for transients is defined as \cite{martinez_2022}:
\begin{equation}
  N_\sigma = \frac{N_\mathrm{S}}{\sqrt{N_\mathrm{S} + N_\mathrm{B}}},
  \label{eq:snratio}
\end{equation}
where $N_\mathrm{S}$ corresponds to the number of source events and
$N_\mathrm{B}$ the number of background events integrated over the exposure time
interval $\Delta T$. Here, we considered a nominal exposure time $\Delta T$ =
10~s, which is between the duration of short and long GRBs. However, this exposure
time should be optimized depending on the detector background rate to have a
meaningful S/N value for source detection. Here, the total background is
considered from all the background components described in Section~\ref{sec:bkg}.

For the source, we considered a typical GRB spectrum. The GRB spectral
models can be defined by: Band function, exponentially attenuated power-law
function (``Comptonized''), and other spectral models \citep{poolakkil_2021}.
For the current purpose, we considered 2300 GRB spectra from the Fermi-GBM GRB
spectral catalog \citep{poolakkil_2021} in this calculation. 

\begin{figure}
  \centering
  \includegraphics [width=0.45\columnwidth]{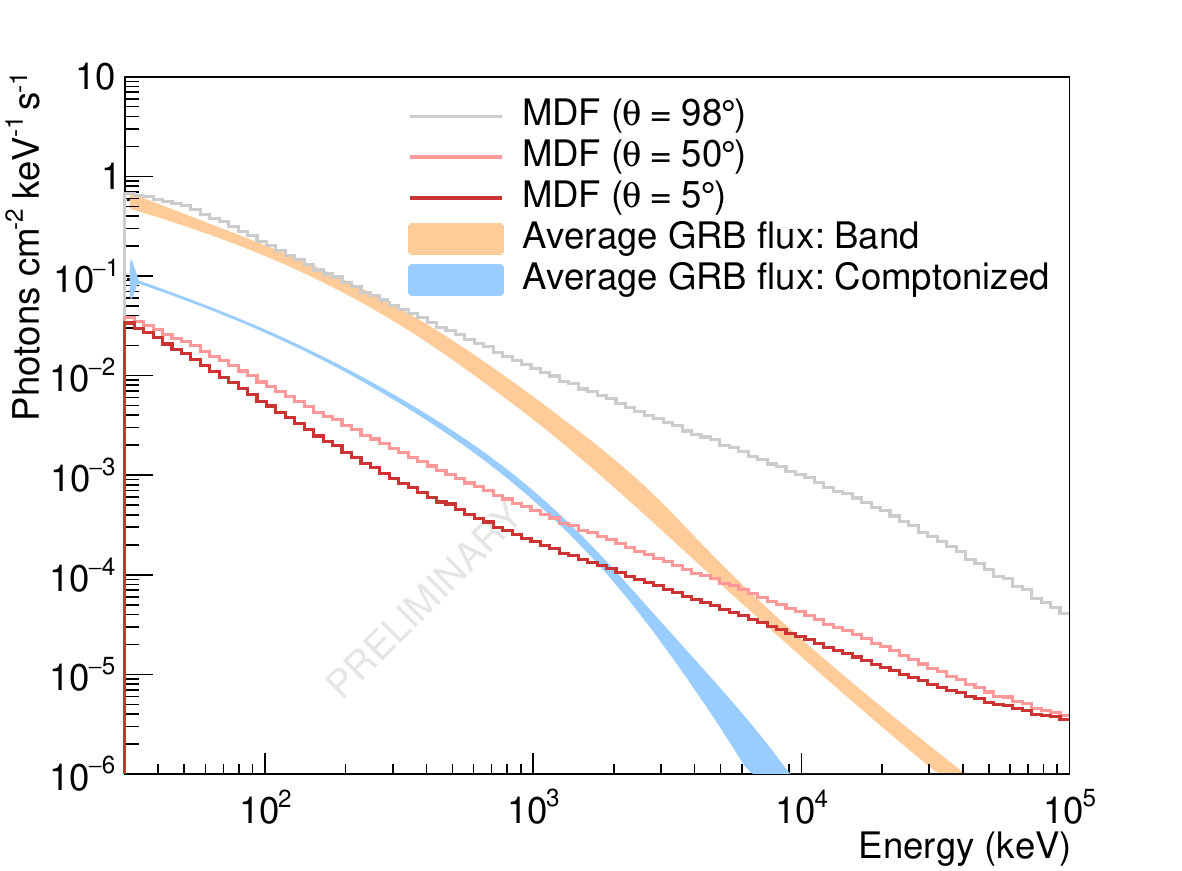}
  \includegraphics [width=0.45\columnwidth]{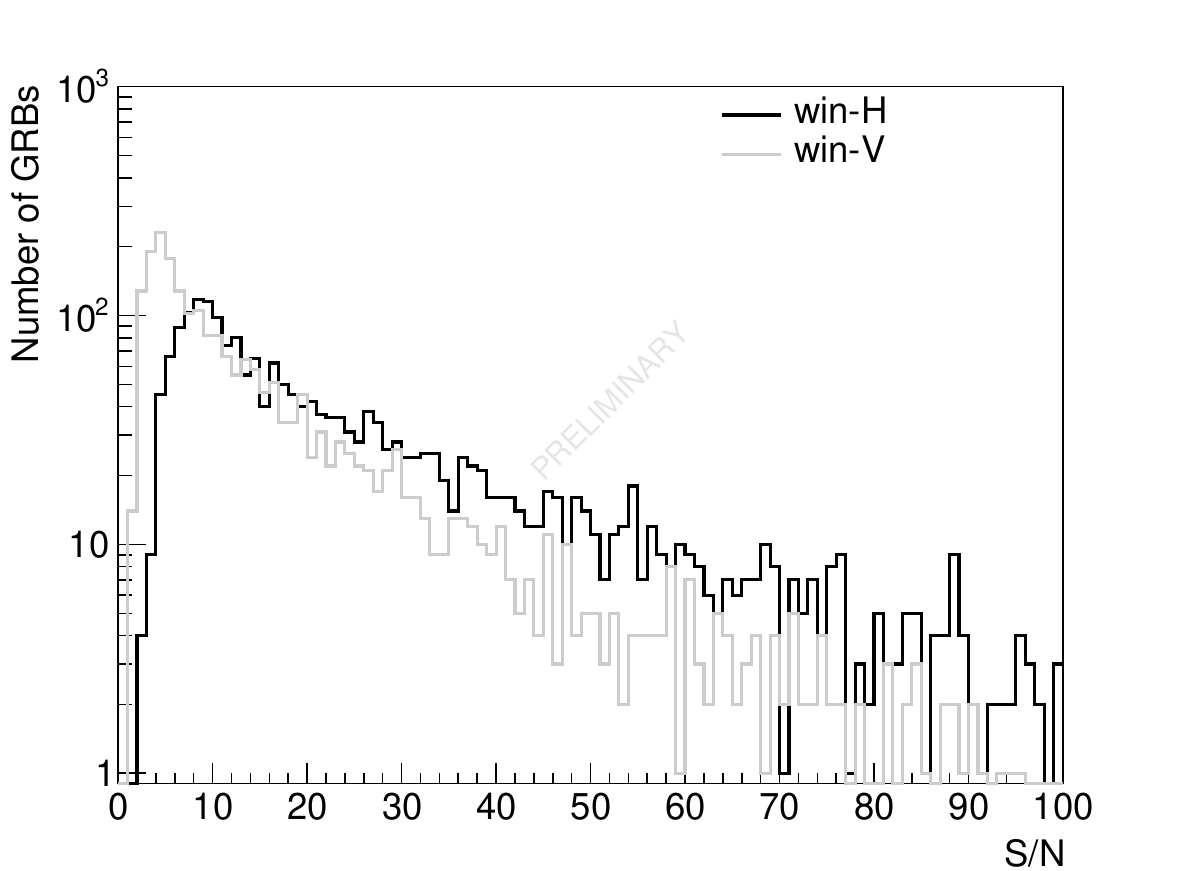}
  \caption{(Left:) Minimum detection fluxes (at different latitudes in the
  orbit) and average GRB fluxes for different best fit models (Band and
  Comptonized) of Fermi-GBM catalog data. (Right:) Signal-to-noise ratio of
  2300 GRBs in Fermi-GBM catalog represented by Comptonized model (for $\Delta$T
  = 10~s and 50$^\circ$ latitude).}
  \label{fig:sens}
\end{figure}

We calculated the minimum detectable flux (MDF) as the notion of sensitivity for
the detector as a function of energy by solving Eq.~\ref{eq:snratio} for
$N_\mathrm{S}$ in each energy bin. The MDF is calculated for a nominal $N_\sigma$
= 3, giving a confidence level of 99.87\%. The values are shown in the left panel
of Fig.~\ref{fig:sens} for the background at three different orbital locations at
5$^\circ$, 50$^\circ$, and 98$^\circ$. The average GRB spectra from the Fermi-GBM
catalog, best fitted by the Band function (which usually represents GRB flux
emission better) and Comptonized functions (gives little restrictive
representation of the flux), are also shown on the same plot by a band of
1$\sigma$ standard error. 

The S/N for all GRBs detected by Fermi-GBM is calculated in the energy range of
30~keV to 100~MeV for an exposure time of 10~s. The S/N values for source
detection in both CALOg windows are shown in the right panel of
Fig.~\ref{fig:sens} for the detector orbital location at 50$^\circ$ latitude.

\begin{figure}
  \centering
  \includegraphics [width=0.45\columnwidth]{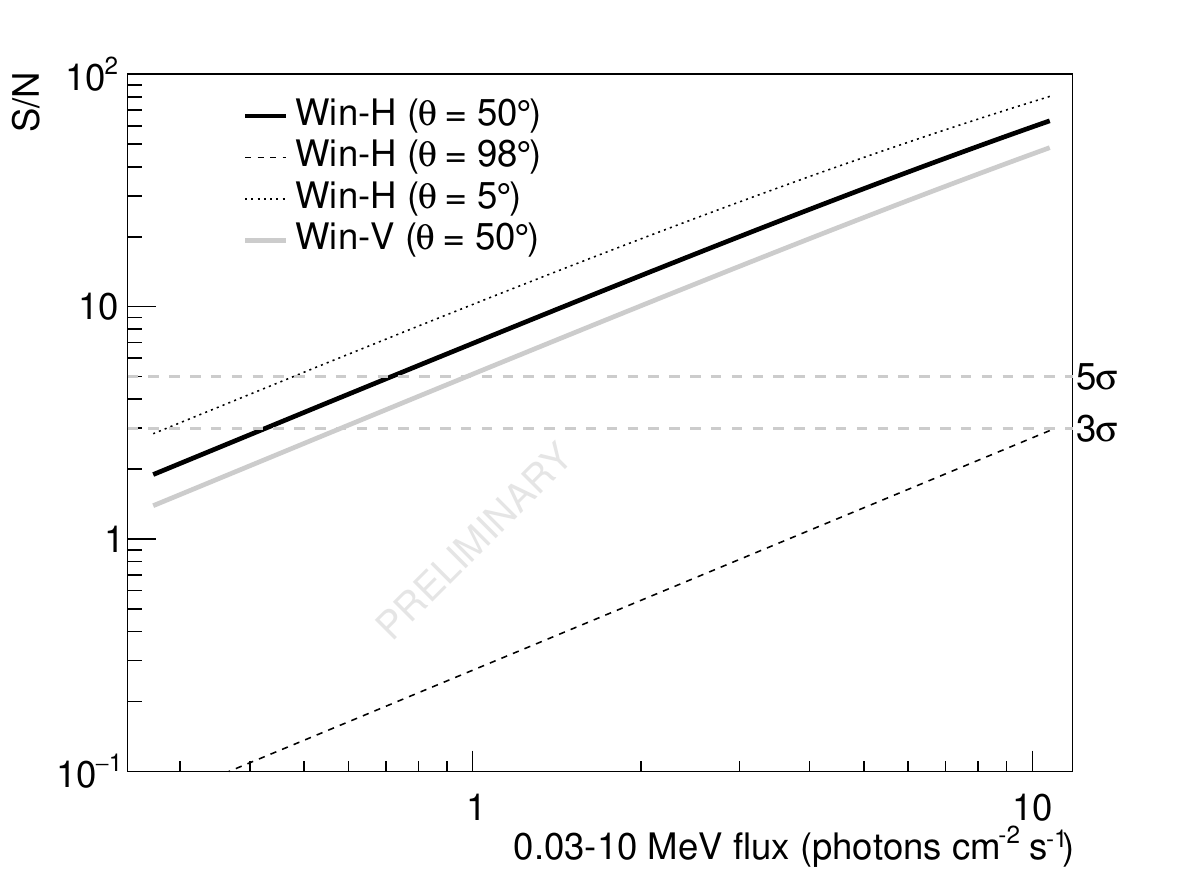}
  \includegraphics [width=0.45\columnwidth]{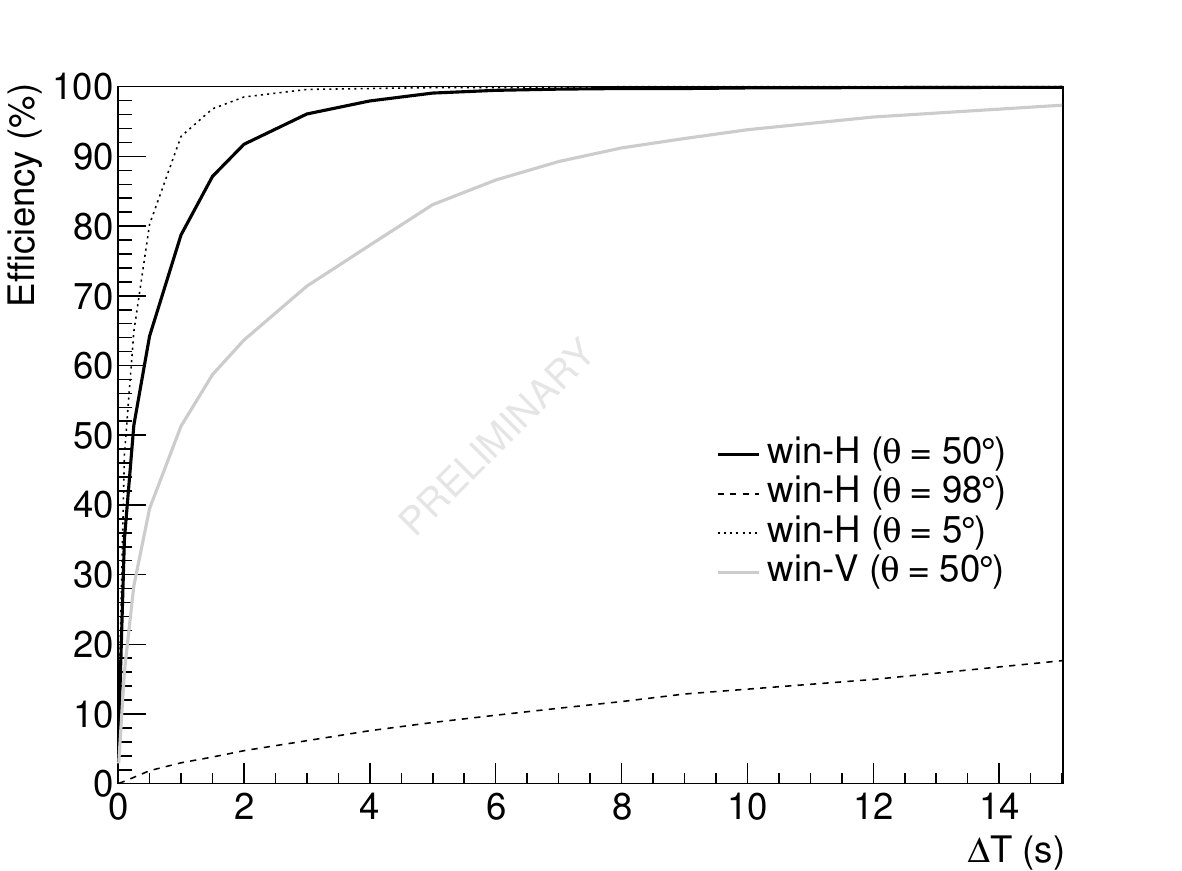}
  \caption{(Left:) Signal-to-noise ratio vs. 0.03--10~MeV flux of an average
  incident GRB model at different latitude positions ($\Delta$T = 10~s).
  (Right:) Efficiency of the GRB detection (number of detection with S/N $>$ 3
  out of 2300 GRBs) vs. exposure time.}
  \label{fig:snreff}
\end{figure}

In order to have an estimation of flux dependence of the GRB detection in terms
of S/N, we considered an average GRB spectral model described by the Comptonized
function. This average model is calculated by the spectral fitting of the average
of all the GRB (Comptonized) spectra from the Fermi-GBM catalog. We varied the
amplitude value of the average spectrum keeping the shape fixed. Then we
calculate the integrated flux over the 30~keV to 10~MeV energy range and the
corresponding S/N (in the 30~keV to 100~MeV energy range and for 10~s exposure
time). The left panel of Fig.~\ref{fig:snreff} shows the integrated flux vs. S/N
plots for both horizontal (for three orbital locations) and vertical windows,
whereas the source is located on-axis to the corresponding windows.

The efficiency of source detection in the detector above some definitive S/N
value depends on the exposure time. To have an estimate of the exposure times
needed for an efficient source detection, we calculate the detection efficiency
varying the exposure times, for both windows and at different orbital locations.
The result is shown in the right panel of Fig.~\ref{fig:snreff}. The calculation
indicates the need for an exposure time of at least about 6~s, for an overall good source detection efficiency.

\section{Timing analysis}
\label{sec:trig}

\begin{figure}
\centering
\begin{minipage}[c]{0.31\textwidth}
\centering
    \includegraphics[width=\textwidth]{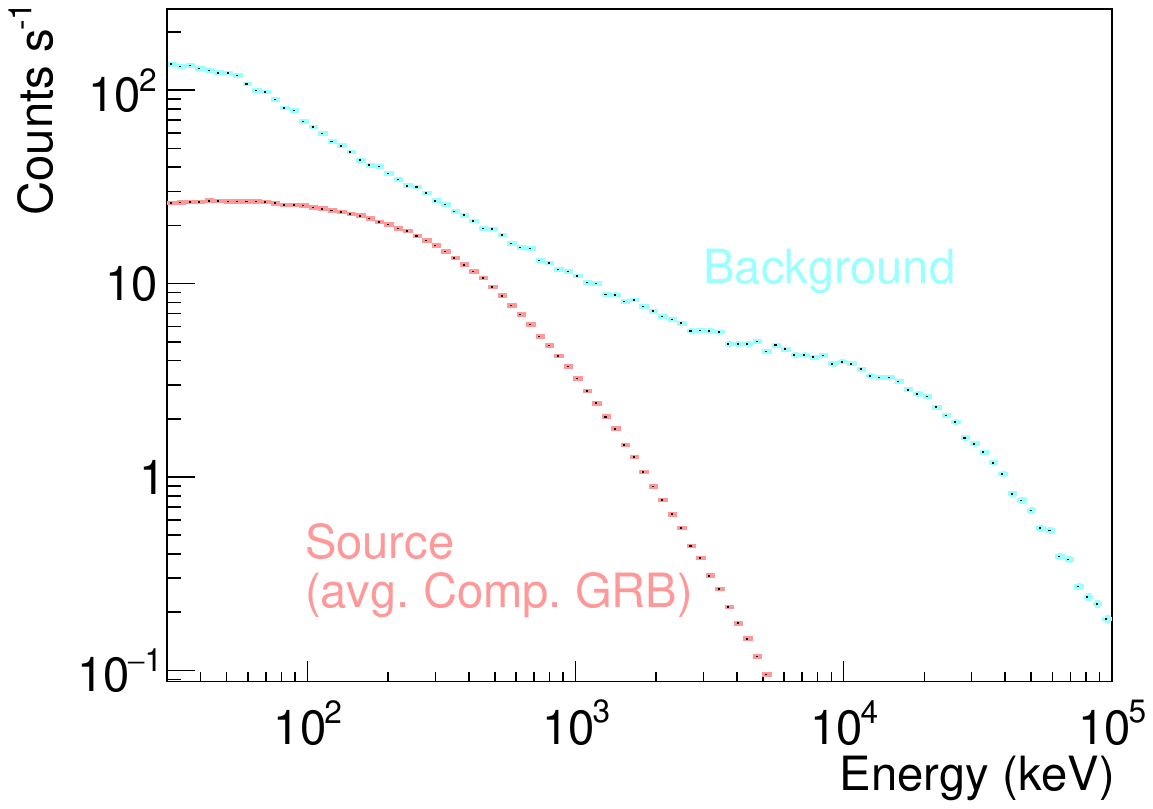}
    \includegraphics[width=\textwidth]{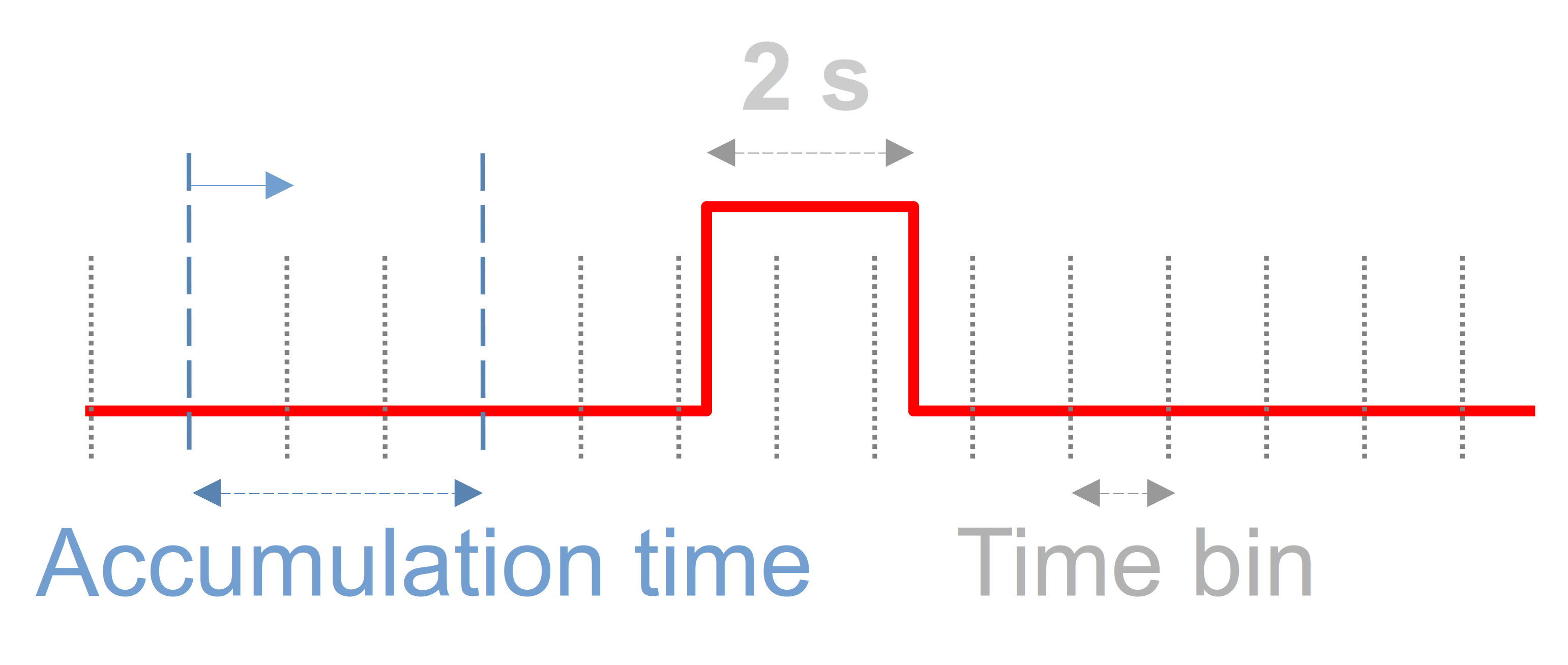}
\end{minipage}
\begin{minipage}[c]{0.5\textwidth}
\centering
    \includegraphics[width=\textwidth]{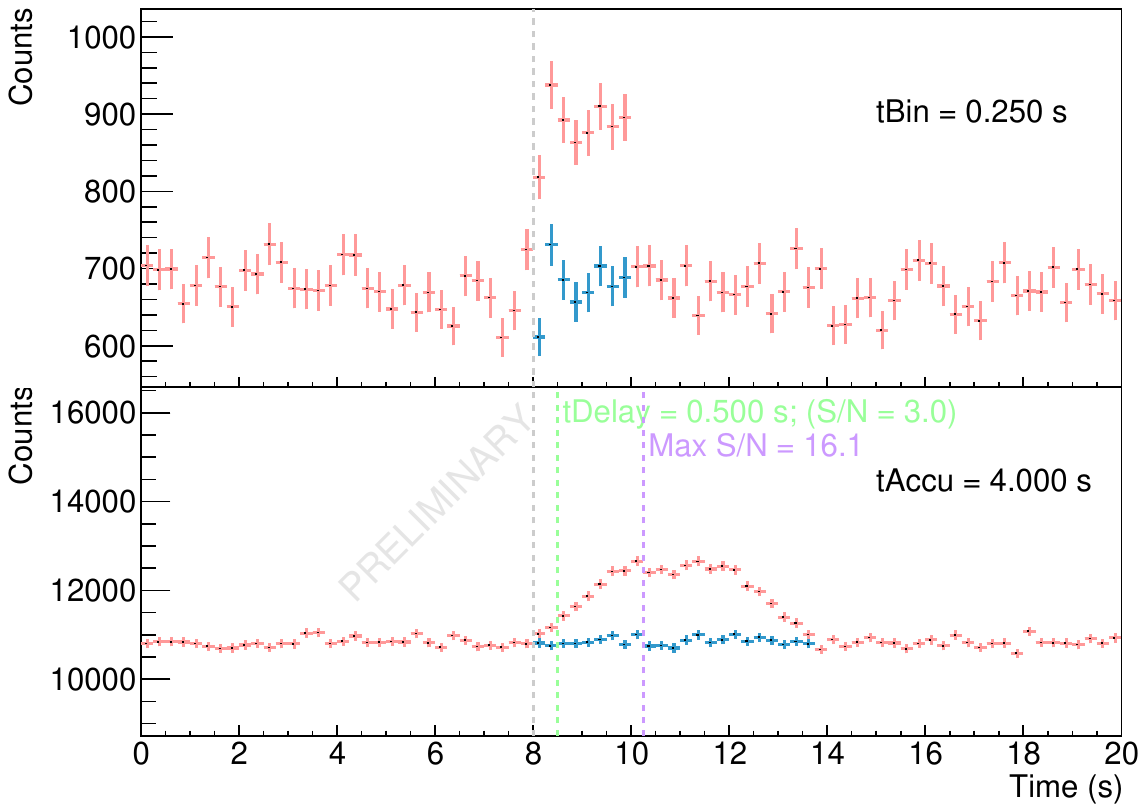}
\end{minipage}
    \caption{(Left top:) Nominal background and source flux distributions
    used for the transient timing analysis. (Left bottom:) Time profile
    of the dummy source (a step function) used for the timing analysis.
    (Right:) Background (blue) and total (source and background shown in red)
    count distribution with time for time bin (tBin) of 250~ms (in upper
    panel), while the same for accumulation time (tAccu) of 4~s is shown in
    the lower panel. Triggering point of the source detection with time delay
    (tDelay) and point of maximum $S/N$ are also denoted. The gray dashed line
    marks the starting point of the source flare.}
    \label{fig:trig}
\end{figure}

Finally, we evaluate the time evolution of the GRB detection, to obtain some
useful information for the design of the GRB trigger of the Zirè CALOg. In this
calculation, we used the simulated background data at 50$^\circ$ latitude and an
average GRB flux (with Comptonized spectral model), as shown in the upper left
panel of Fig.~\ref{fig:trig}. We considered a simple time profile for the
transient source --- a step function with a nominal 2~s outburst time, as shown
in the lower left panel of Fig.~\ref{fig:trig}. We collected the source and
background counts in a time bin (tBin) of 250~ms to reproduce the time series
data. The time counts in the detector, only for the background and also for the
total contribution along with the source, are shown in the upper panel of
Fig.~\ref{fig:trig}. To calculate the existence of a source in the detector
acceptance, we compared the total detector counts in a certain accumulation time
(tAccu) with the background counts. The S/N is calculated in each of the
accumulation time bin. The corresponding counts with tAccu = 4~s are
shown in the bottom right panel of the same figure. The time delay (tDelay) for
a 3$\sigma$ confidence of source detection is calculated and indicated in the
figure. The information from this calculation can be used to design an onboard
trigger algorithm for transient source detection and will be reported in the
future.

\section{Conclusions and outlook}
\label{sec:conc}

The Zirè detector onboard the NUSES satellite will contribute not only to the
investigation of scientific topics involving low-energy solar and Galactic CR
and $\gamma$ rays in the keV--MeV regime, but it will also play a crucial role
in the development and testing of advanced technologies, such as the SiPM-based
readout system to be used for detectors in space. The CALOg instrument can
provide good transient source detection in its complementary role along with the
detection of CR particles. Under the diverse background radiation environment the
satellite will encounter along its orbit, the detector can provide good transient
sensitivity for a large fraction of its live time away from the polar (and
south-Atlantic anomaly) regions. More detailed calculation of transient source
triggering efficiency in the detector is undergoing, also considering the use of
the convolutional neural network for this purpose and will be communicated in
future.

\section*{Acknowledgements}
\label{ackn}

NUSES is funded by the Italian Government (CIPE n. 20/2019), by the Italian Ministry of Economic Development (MISE reg. CC n. 769/2020), by the Italian Space Agency (CDA ASI n. 15/2022), by the European Union NextGenerationEU under the MUR National Innovation Ecosystem grant ECS00000041 - VITALITY - CUP D13C21000430001 and by the Swiss National Foundation (SNF grant n. 178918). This study was carried out also in collaboration with the Ministry of University and Research, MUR, under contract n. 2024-5-E.0 - CUP n. I53D24000060005. This research, leading to the beam test results, also received partial funding from the European Union’s Horizon Europe research and innovation programme under grant agreement No. 101057511.

\bibliographystyle{JHEP}
\bibliography{references}

\clearpage

\begin{center}
\textbf{The NUSES Collaboration}\\[18pt]

M.~Abdullahi$^{a,b}$, R.~Aloisio$^{a,b}$, F.~Arneodo$^{c,d}$, S.~Ashurov$^{a,b}$, U.~Atalay$^{a,b}$, F.~C.~T.~Barbato$^{a,b}$, R.~Battiston$^{e,f}$, M.~Bertaina$^{g,h}$, E.~Bissaldi$^{i,j}$, D.~Boncioli$^{k,b}$, L.~Burmistrov$^{l}$, F.~Cadoux$^{l}$, I.~Cagnoli$^{a,b}$, E.~Casilli$^{a,b}$, D.~Cortis$^{b}$, A.~Cummings$^{m}$, M.~D'Arco$^{l}$, S.~Davarpanah$^{l}$, I.~De~Mitri$^{a,b}$, G.~De~Robertis$^{i}$, A.~Di~Giovanni$^{a,b}$, A.~Di~Salvo$^{h}$, L.~Di~Venere$^{i}$, J.~Eser$^{n}$, Y.~Favre$^{l}$, S.~Fogliacco$^{a,b}$, G.~Fontanella$^{a,b}$, P.~Fusco$^{i,j}$, S.~Garbolino$^{h}$, F.~Gargano$^{i}$, M.~Giliberti$^{i,j}$, F.~Guarino$^{o,p}$, M.~Heller$^{l}$, T.~Ibrayev$^{c,d,q}$, R.~Iuppa$^{e,f}$, A.~Knyazev$^{c,d}$, J.~F.~Krizmanic$^{r}$, D.~Kyratzis$^{a,b}$, F.~Licciulli$^{i}$, A.~Liguori$^{i,j}$, F.~Loparco$^{i,j}$, L.~Lorusso$^{i,j}$, M.~Mariotti$^{s,t}$, M.~N.~Mazziotta$^{i}$, M.~Mese$^{o,p}$, M.~Mignone$^{g,h}$, T.~Montaruli$^{l}$, R.~Nicolaidis$^{e,f}$, F.~Nozzoli$^{e,f}$, A.~Olinto$^{u}$, D.~Orlandi$^{b}$, G.~Osteria$^{o}$, P.~A.~Palmieri$^{g,h}$, B.~Panico$^{o,p}$, G.~Panzarini$^{i,j}$, D.~Pattanaik$^{a,b}$, L.~Perrone$^{v,w}$, H.~Pessoa~Lima$^{a,b}$, R.~Pillera$^{i,j}$, R.~Rando$^{s,t}$, A.~Rivetti$^{h}$, V.~Rizi$^{k,b}$, A.~Roy$^{a,b}$, F.~Salamida$^{k,b}$, R.~Sarkar$^{a,b}$, P.~Savina$^{a,b}$, V.~Scherini$^{v,w}$, V.~Scotti$^{o,p}$, D.~Serini$^{i}$, D.~Shledewitz$^{e,f}$, I.~Siddique$^{a,b}$, L.~Silveri$^{c,d}$, A.~Smirnov$^{a,b}$, R.~A.~Torres~Saavedra$^{a,b}$, C.~Trimarelli$^{a,b}$, P.~Zuccon$^{e,f}$, S.~C.~Zugravel$^{h}$.

 \vspace{5ex}

\begin{tabular}{c}
$^{a}$ Gran Sasso Science Institute (GSSI);\\
$^{b}$ Istituto Nazionale di Fisica Nucleare (INFN) - Laboratori Nazionali del Gran Sasso;\\
$^{c}$ Center for Astrophysics and Space Science (CASS);\\
$^{d}$ New York University Abu Dhabi, UAE;\\
$^{e}$ Dipartimento di Fisica - Università di Trento;\\
$^{f}$ Istituto Nazionale di Fisica Nucleare (INFN) - Sezione di Trento;\\
$^{g}$ Dipartimento di Fisica - Università di Torino;\\
$^{h}$ Istituto Nazionale di Fisica Nucleare (INFN) - Sezione di Torino;\\
$^{i}$ Istituto Nazionale di Fisica Nucleare (INFN) - Sezione di Bari;\\
$^{j}$ Dipartimento di Fisica M. Merlin dell’ Università e del Politecnico di Bari;\\
$^{k}$ Dipartimento di Scienze Fisiche e Chimiche -Università degli Studi di L’Aquila;\\
$^{l}$ Départment de Physique Nuclèaire et Corpuscolaire - Université de Genève, Faculté de Science;\\
$^{m}$ Department of Physics and Astronomy and Astrophysics, Institute for Gravitation and the Cosmos;\\
$^{n}$ Department of Astronomy and Astrophysics, University of Columbia;\\
$^{o}$ Istituto Nazionale di Fisica Nucleare (INFN) - Sezione di Napoli;\\
$^{p}$ Dipartimento di Fisica E. Pancini - Università di Napoli Federico II;\\
$^{q}$ now at The School of Physics, The University of Sydney;\\
$^{r}$ CRESST/NASA Goddard Space Flight Center;\\
$^{s}$ Dipartimento di Fisica e Astronomia - Università di Padova;\\
$^{t}$ Istituto Nazionale di Fisica Nucleare (INFN) - Sezione di Padova;\\
$^{u}$ Columbia University, Columbia Astrophysics Laboratory;\\
$^{v}$ Dipartimento di Matematica e Fisica “E. De Giorgi” - Università del Salento;\\
$^{w}$ Istituto Nazionale di Fisica Nucleare (INFN) - Sezione di Lecce. \\
\end{tabular}

\end{center}

\end{document}